\documentclass[12pt]{iopart}
\usepackage{graphicx}
\usepackage{cite}
\newcommand{\eq}[1]{$ #1 $}
\newcommand{\eqw}[2]{$ #1~{\rm #2} $}

\def\etal{{\it et al.}}

\def\vi{V_{ie1}}
\def\vii{V_{ie2}}
\def\vhi{V_{H1}}
\def\vhii{V_{H2}}

\def\nmid{\tilde{I}_{D} / \bar{I}_{D}}

\begin{document}

\title[{\scriptsize Electrostatic Ion-Cyclotron Instabilities Modified by the Parallel and Perpendicular Plasma Flow Velocity Shears}]{\large Electrostatic Ion-Cyclotron Instabilities Modified by the Parallel and Perpendicular Plasma Flow Velocity Shears}

\author{T. Kaneko
\footnote[3]{kaneko@ecei.tohoku.ac.jp}
, H. Saito, H. Tsunoyama, and R. Hatakeyama}

\address{Department of Electronic Engineering, Tohoku University, Sendai 980-8579, Japan}


\begin{abstract}
\baselineskip 13pt
The external and independent control of plasma flow velocity shears parallel and perpendicular to magnetic field lines is realized using segmented collisionless-plasma sources.
Electrostatic ion-cyclotron instabilities are observed to be suppressed by the perpendicular flow velocity shears, the suppression physics of which is found to be distinguished into two aspects. On the other hand, the parallel flow velocity shears are demonstrated to destabilize the ion-cyclotron instabilities depending on the sign of the parallel shear in the absence of field-aligned electron drift flow.

\end{abstract}

\baselineskip 14pt
\section{Introduction}
\hspace{4ex}
Investigations of magnetic-field-aligned (parallel) and transverse (perpendicular) sheared plasma flows have been widely performed in connection with the excitation or the suppression of plasma fluctuations and turbulences, which are considered to generate energetic particles and induce cross-field transport in fusion-oriented and space plasmas.
In basic laboratory experiments, a number of works on each of the flow shears have been reported.
In the parallel shear case, some experimental investigations related to parallel-flow-shear driven instabilities, such as the D'Angelo mode \cite{dangelo309}, ion-acoustic \cite{agrimson5282,teodorescu185003}, drift-wave \cite{kaneko125001}, and ion-cyclotron \cite{agrimson260,teodorescu105001} instabilities, have been performed in various situations.
In the perpendicular shear case, on the other hand, many experimental investigations on the relation between the  perpendicular flow shears and the instabilities, such as Kelvin-Helmholtz (KH) \cite{jassby1567}, inhomogeneous energy density driven (IEDD) \cite{koepke3355,amatucci1978}, drift-mode \cite{mase2281,yoshinuma191}, and flute-mode \cite{komori210} instabilities, have been performed. 

Although an instability in the ion-cyclotron range of frequencies, which plays an important role in heating of ions, plasma cross-field diffusion, and anomalous resistivity in space plasmas, has been investigated for the case of the IEDD instability and is reported to be excited by the perpendicular shear, this IEDD instability is treated as different from the conventional ion-cyclotron instability \cite{koepke3355}.
Originally, the ion-cyclotron instability was reported to be driven by an electron current generated along magnetic-field lines in a plasma column with a positively biased small electrode immersed in the plasma, which is called as the current-driven ion cyclotron instability \cite{drummond1507,dangelo633,correll1800}.
On the other hand, Hatakeyama {\etal} pointed out an importance of two-dimensional potential structure near the small electrode \cite{hatakeyama28,sato1661,hatakeyama285}, which gives rise to a two-dimensional relaxation oscillation of ions in a magnetized plasma.
This relaxation oscillation is called as ``potential-driven ion-cyclotron oscillation" and is distinguished from the above ``current-driven instability".

Since the effects of the flow velocity shears on these two types of ion-cyclotron instabilities have not been investigated yet, it is significant to clarify these effects for the understanding of dynamical ion processes in the space plasmas.
Thus, we here focus upon the characteristics of the ion-cyclotron instabilities modified by the parallel and perpendicular flow velocity shears which are actively controlled using our newly developed plasma source.
\par

\section{Experimental Setup}

\begin{figure}[b]
\vspace{5mm}
\begin{center}
  	\includegraphics[width=8.5cm]{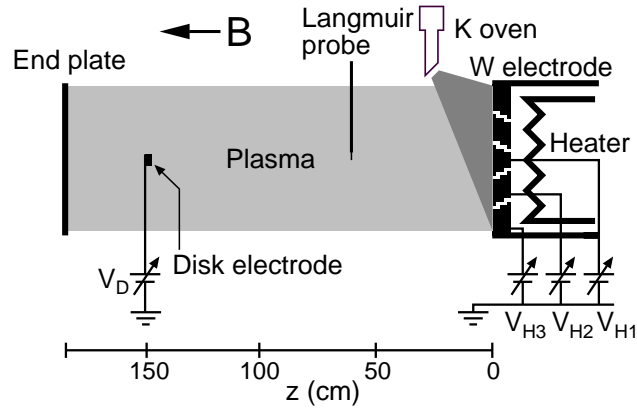}\\
\caption{Schematic of experimental setup of perpendicular flow velocity shear. \hfill}
\label{fig:perp-s}
\end{center}
\end{figure}

\hspace{4ex}
Experiments are performed in the $\rm Q_T-$Upgrade machine of Tohoku university. In the case of perpendicular flow velocity shear experiments, a plasma is produced by the surface ionization of potassium atoms on a 10.0-cm-diameter tungsten (W) hot plate under a magnetic field of \eqw{B=3}{kG} in a single-ended Q machine as shown in Fig.~\ref{fig:perp-s}. The hot plate is concentrically segmented into three section, each of which is electrically isolated and is individually biased. Thus, the radially-different plasma potential, i.e., radial electric field is generated even in the fully-ionized collisionless plasma.
Hereinafter, the hot-plate electrodes set in order from the center to the outside are called as the first, second, and third electrodes and voltages applied to them are defined as \eq{V_{H1}, V_{H2}}, and \eq{V_{H3}}, respectively. \eq{V_{H3}} is kept at 0 V throughout this paper.
A small radially movable Langmuir probe is used to measure radial profiles of plasma parameters. Here, the axial position \eq{z} is defined as the distance from the hot plate (\eqw{z=0}{cm}) toward a glass end plate located at \eqw{z=170}{cm}, which terminates the plasma column.
The plasma is produced almost within the second electrode and the plasma density \eq{n_p} is about \eqw{10^9}{cm^{-3}} at the radial center, gradually decreasing toward the outside. Electron \eq{T_e} and ion \eq{T_i} temperatures are around \eqw{0.2}{eV} and their profiles are almost uniform in the radial direction.
In order to investigate effects of the perpendicular shear on the instabilities in the ion-cyclotron frequency range, a small disk electrode, the diameter of which is 2 mm, is inserted at the center of the plasma cross section around \eqw{z=150}{cm} (see Fig.~\ref{fig:perp-s}), which has been known to excite the ion-cyclotron instabilities by changing a bias voltage \eq{V_D} applied to the electrode \cite{dangelo633,correll1800,hatakeyama28,sato1661,hatakeyama285}.

In the case of parallel flow shear experiments, on the other hand, a plasma is produced by a modified plasma synthesis method, where potassium ion and electron emitters are oppositely set at cylindrical machine ends under a strong magnetic field of \eqw{B= 1.6}{kG} \cite{kaneko4218}.
The ion emitter is the same as the hot plate in the perpendicular flow shear experiment and the electron emitter using a 10.8-cm-diameter nickel plate coated with barium oxide (BaO) is additionally mounted at a distance of 170 cm from the ion emitter, which is the principal feature in the parallel shear experiment.
A negatively biased stainless (Sus) grid, the voltage of which is typically \eqw{V_g= -60}{V}, is installed at a distance of 10 cm from the ion emitter surface.  Since the grid reflects the electrons flowing from the electron emitter, an electron velocity distribution function parallel to the magnetic field is considered to become Maxwellian, namely there is no electron drift flow.
In this synthesized plasma, the electron emitter is negatively biased at typically \eqw{V_{ee} \simeq -4.0}{V}, which determines the plasma potential \eq{\phi}, and thus, a voltage applied to the ion emitter can control the potential difference between the plasma and the ion emitter. This potential difference can accelerate the ions and generate the field-aligned ion flow. Since the ion emitter is concentrically segmented into three sections, each of which is electrically isolated and is individually biased (\eq{\vi, \vii, V_{ie3}}) as mentioned above, the field-aligned ion flows with radially-different energies, i.e., ion flow velocity shears, are expected to be generated in the radially-uniform plasma potential \cite{kaneko4218}.

\section{Experimental Results and Discussion}
\subsection{Perpendicular flow velocity shear}

\begin{figure}[h]
\begin{center}
  	\includegraphics[width=5.7cm]{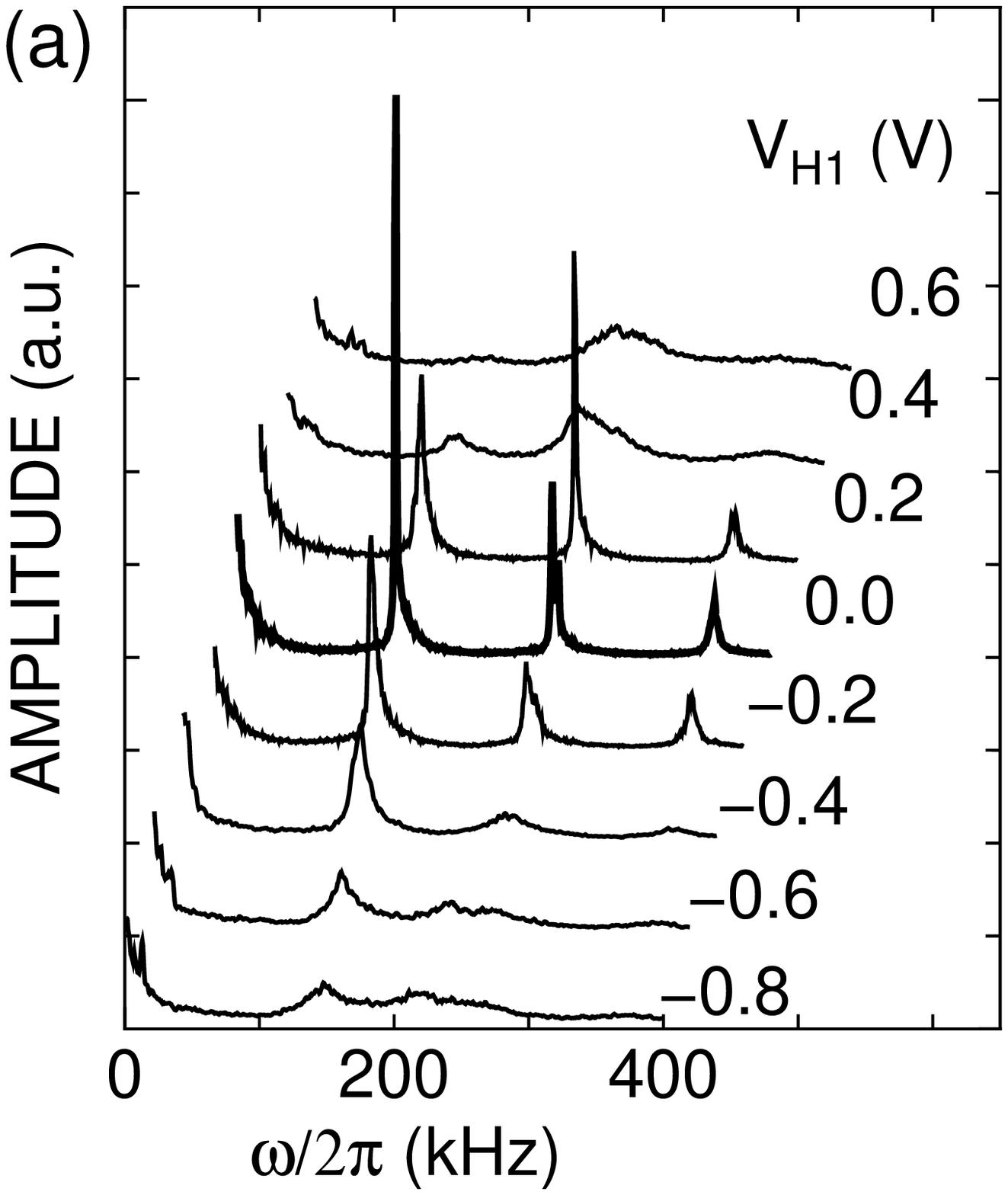}
  	\hspace*{5mm}
  	\includegraphics[width=6.2cm]{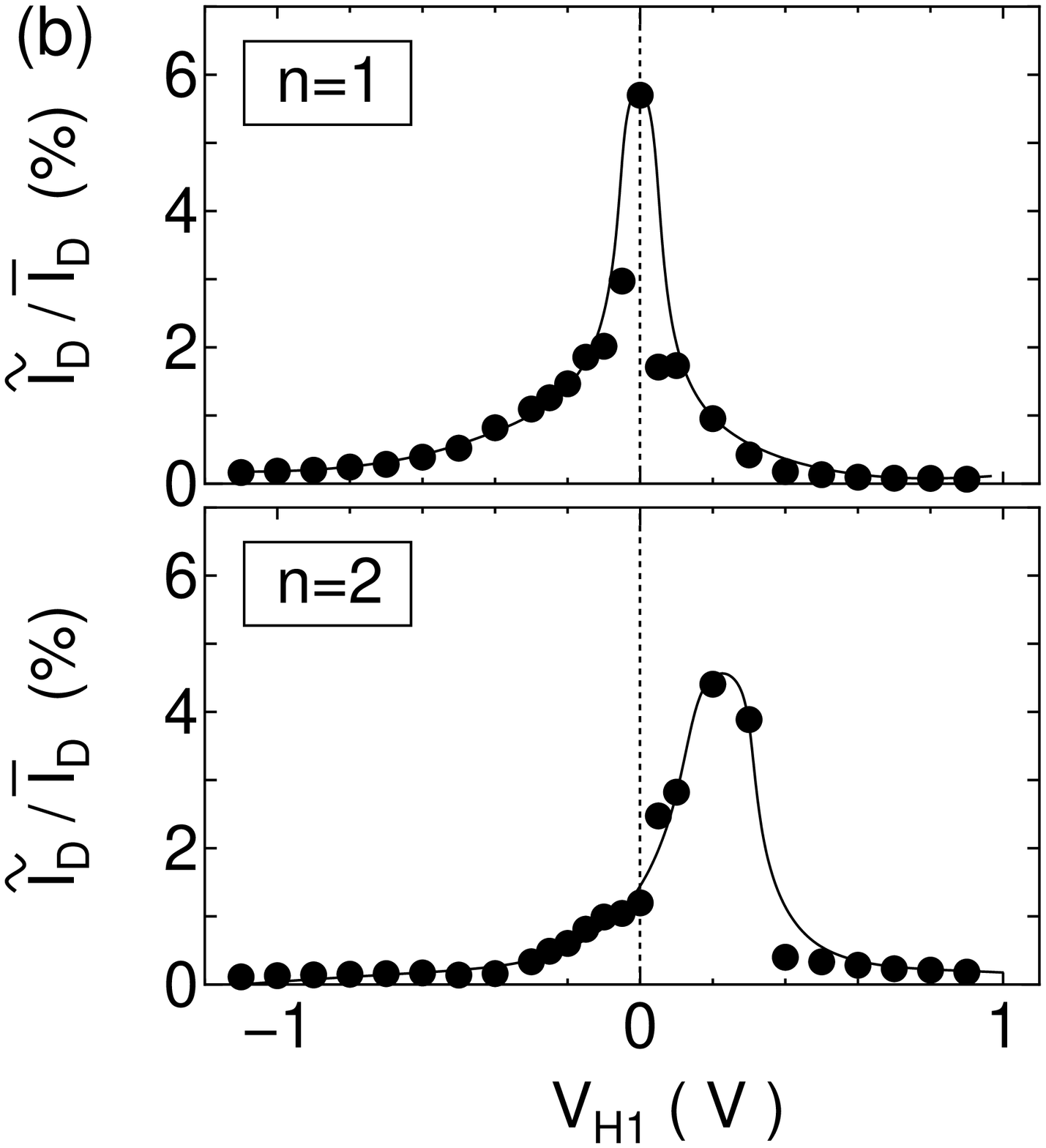}\\
\caption{(a) Frequency spectra of electron saturation current and (b) normalized fluctuation amplitudes \eq{\nmid} as a function of \eq{\vhi} for \eqw{\vhii =0}{V} and \eqw{V_D =90}{V} at \eqw{r= 0}{cm} in the perpendicular shear case.}
\label{fig:vh1-amp-90}
\end{center}
\end{figure}

\hspace{4ex}
First, we investigate the effects of the perpendicular flow velocity shear on the ion-cyclotron instability, which is excited by the positively biased (\eqw{V_D =90}{V}) small disk electrode. Figure~\ref{fig:vh1-amp-90}(a) shows frequency spectra of an electron current \eq{I_D} flowing to the small disk electrode as a function of \eq{\vhi} for \eqw{\vhii =0}{V} at \eqw{r= 0}{cm}.
When \eq{I_D} is increased by changing \eq{V_D}, fluctuations in the regions of ion-cyclotron frequency (\eqw{\omega_{ci}/2\pi =116}{kHz} at \eqw{B=3}{kG}) and its multiplied frequencies are excited.
It is confirmed that the fluctuation frequencies \eq{\omega / 2\pi} shift with an increase in magnetic-field strength and correspond to the ion-cyclotron and harmonic frequencies (\eq{\omega \simeq n\omega_{ci}}, \eq{n=1, 2, 3, \cdot \cdot \cdot}).
These ion-cyclotron instabilities are excited in the absence of the perpendicular shear for \eqw{\vhi = 0}{V}, and are gradually suppressed with an increase or a decrease in \eq{\vhi}.

Normalized fluctuation amplitudes \eq{\nmid} of the electron current for \eqw{V_D =90}{V} as a function of \eq{\vhi} for \eqw{\vhii =0}{V} is presented in Fig.~\ref{fig:vh1-amp-90}(b). 
In the case of the doubled-frequency mode (n=2) with \eqw{\omega / 2\pi \simeq 240}{kHz}, the fluctuation amplitude \eq{\nmid} increases with an increase in \eq{\vhi} and has the maximum value for \eqw{\vhi \simeq +0.2}{V}, where the generated hill-shaped radial potential profile forms the slight perpendicular flow velocity shear. This fluctuation is found to be gradually stabilized with an increase or a decrease in \eq{\vhi} which starts from \eqw{+0.2}{V}, and is almost suppressed for \eqw{\vhi = \pm 1}{V}.
Gavrishchaka {\etal} reported that the theoretical growth rate increases with an increase in the perpendicular flow velocity shear and gradually decrease with a further increase in the shear \cite{gavri3091}.
The ion-cyclotron instability treated in the theory of Gavrishchaka {\etal} is concluded to be destabilized by the combination of a magnetic-field-aligned electron current and a localized transverse electric field, i.e., perpendicular velocity shear, which suggests that the current-driven type instability is enhanced by the slight velocity shear and suppressed by the large velocity shear.
Since our experimental results in the case of n=2 mode clearly show the same tendency as the above theoretical growth rate, the ion-cyclotron instability excited by the weak velocity shear is identified as the current-driven type instability and is found to be suppressed by the large velocity shear.

In the case of the fundamental ion-cyclotron mode (n=1) with \eqw{\omega / 2\pi \simeq 120}{kHz}, on the other hand, the ion-cyclotron instability is excited and has the maximum fluctuation amplitude for \eqw{\vhi = 0}{V}, which signifies the flat radial potential, i.e., the absence of the velocity shear. This instability is gradually suppressed with an increase or a decrease in \eq{\vhi}, i.e., the shear suppression is symmetrical with respect to \eqw{\vhi = 0}{V}, which is distinctly different from the case of n=2 mode.
Since the effect of the shear on this instability cannot be explained by the conventional theory for the current-driven type instability, our attention is directed to another concept of ion-cyclotron instability called as the potential-driven type instability \cite{hatakeyama28,sato1661,hatakeyama285}.
This type of instability is caused by not the current along magnetic-field lines but the two-dimensional potential structure just in front of the small disk electrode leading to the synchronized periodic ion gyro-motion. 
When the perpendicular shear is generated, the trajectory of the ion is distorted by the shear and the period of the ion gyro-motion is changed. 
Therefore, the suppression of the potential-driven type ion-cyclotron instability by the perpendicular shear can be understood in terms of the decorrelation of the synchronized periodic ion gyro-motion.
This implies that the potential-driven type instability has the maximum fluctuation amplitude in the absence of the perpendicular flow velocity shear.

\subsection{Parallel flow velocity shear}

\begin{figure}[htbp]
\begin{center}
  	\includegraphics[width=5.7cm]{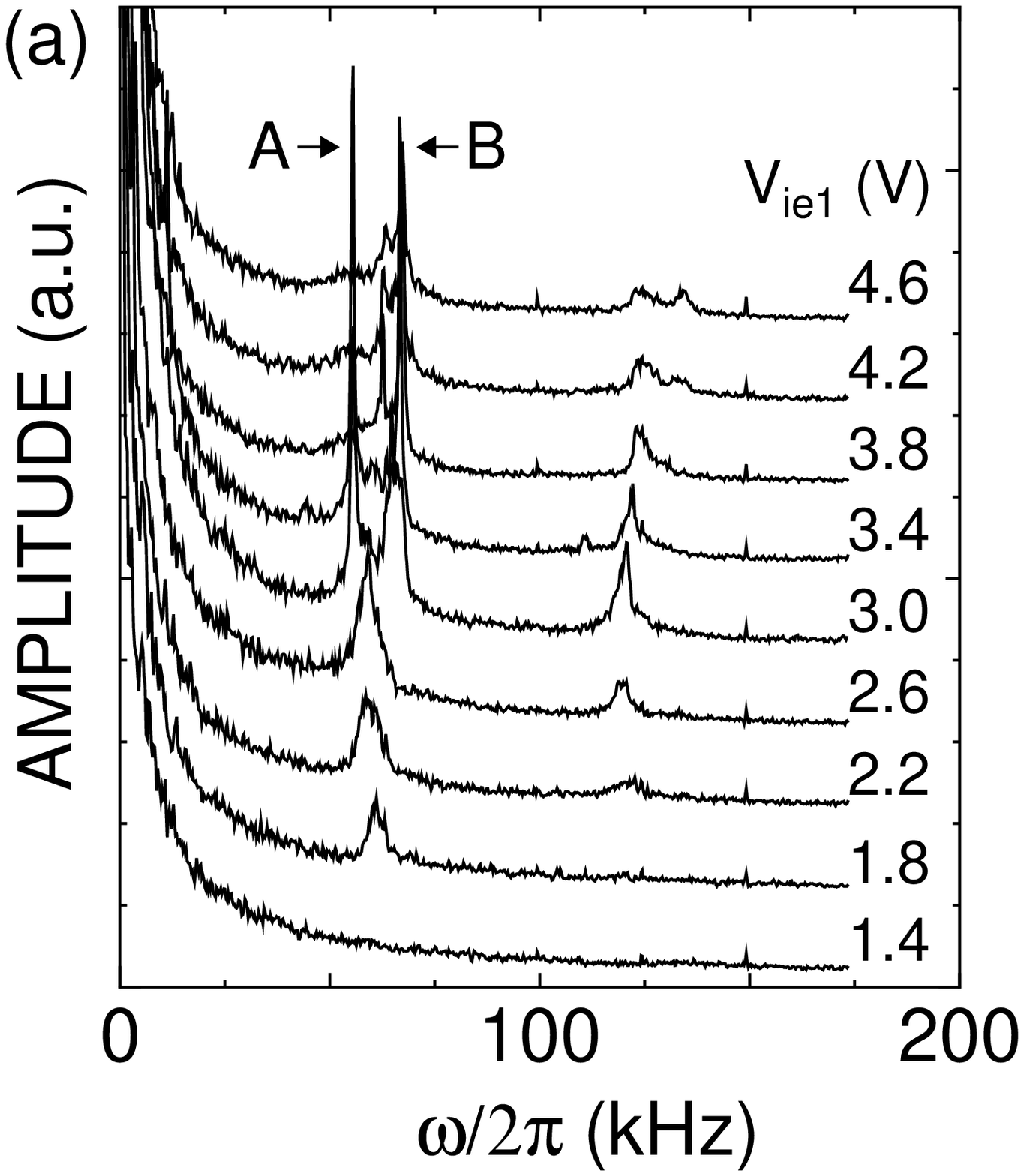}
  	\hspace*{5mm}
  	\includegraphics[width=7cm]{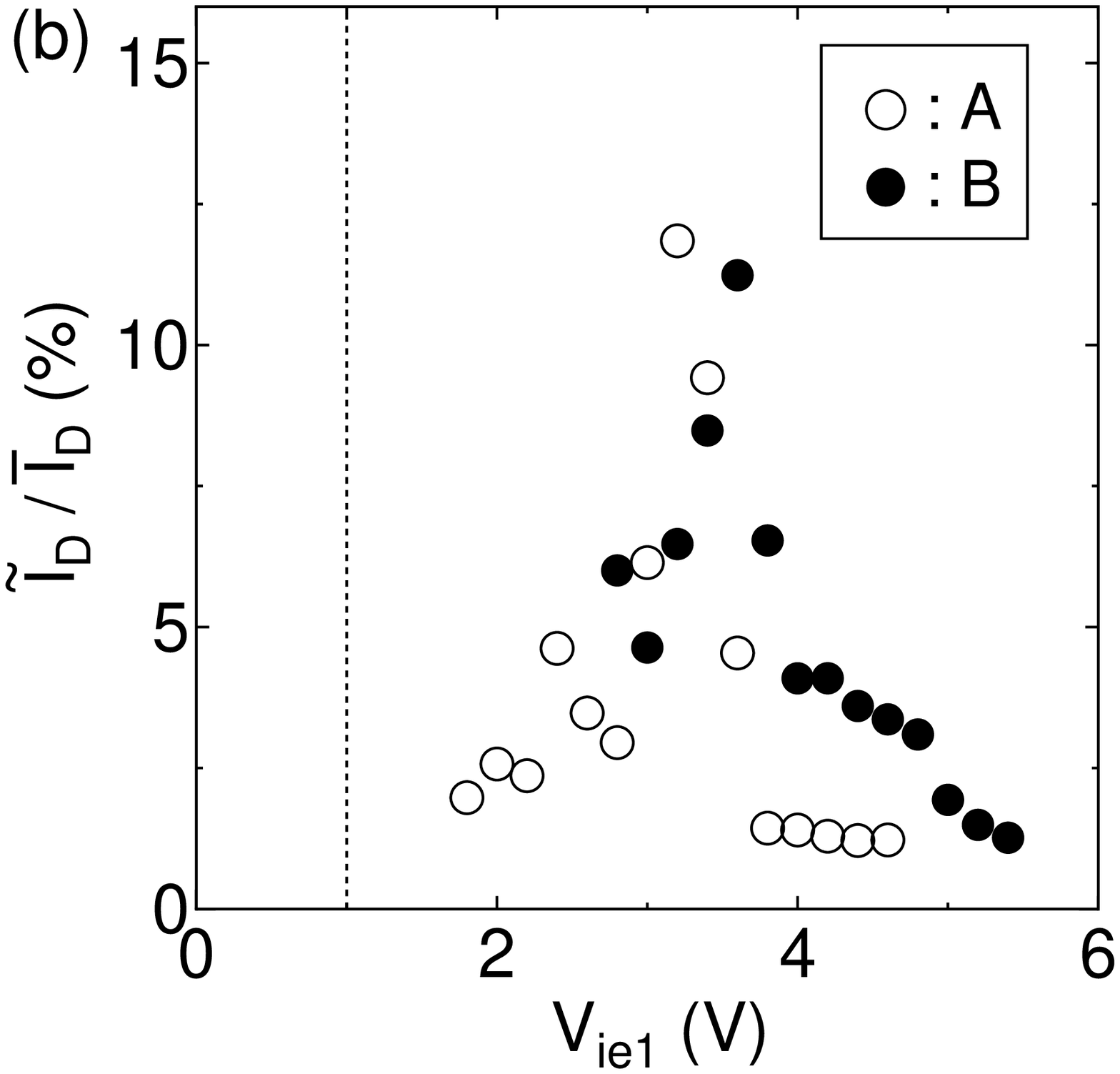}\\
\caption{(a) Frequency spectra of electron saturation current and (b) normalized fluctuation amplitudes \eq{\nmid} as a function of \eq{\vhi} for \eqw{\vhii =0}{V} and \eqw{V_D =90}{V} at \eqw{r= 0}{cm} in the parallel shear case.}
\label{fig:para-amp}
\end{center}
\end{figure}

\hspace*{4ex}
Figure~\ref{fig:para-amp} shows (a) frequency spectra of electron current of the disk electrode and (b) normalized fluctuation amplitudes \eq{\nmid} as a function of \eq{\vi} for \eqw{\vii =1.0}{V} and \eqw{V_D=90}{V} at \eqw{r=0}{cm} in the parallel flow velocity shear case, where a dotted line in Fig.~\ref{fig:para-amp}(b) indicates the value of \eq{\vi} corresponding to the absence of the parallel shear. It is to be noted that the plasma density in the case of parallel shear experiment is about \eqw{10^8}{cm^{-3}} which is a few times smaller than that in the perpendicular shear case. As a result, the ion-cyclotron instability is not excited at \eq{\vi = \vii} even for \eqw{V_D=90}{V}.
When \eq{\vi} is positively biased, however, the fluctuation in the ion-cyclotron frequency range begins to grow at \eqw{\vi \simeq 1.8}{V} (A: open circles) and \eq{\nmid} gradually increases. This \eq{\nmid} has the maximum value at \eqw{\vi \simeq 3.2}{V}, and decreases for larger \eq{\vi}. Since this phenomenon is similar to the dependence of the drift-wave instability on the parallel shear, the ion-cyclotron instability is also enhanced by the parallel shear in the same mechanism as the case of the drift-wave instability as described in Ref.~\cite{kaneko125001}.

From the results of the frequency spectra, the frequencies of the fluctuation are found to be slightly smaller (A: open circles) and larger (B: closed circles) than the ion-cyclotron frequency (\eqw{\simeq 63}{kHz} at \eqw{B=1.6}{kG}). This frequency shift is considered to be caused by the shear effects and is now analyzed by using the kinetic dispersion relation. \\

\section{Conclusion}

\hspace*{4ex}
Our experiments demonstrate that the ion-cyclotron instabilities are suppressed by the perpendicular flow velocity shear.
In addition, it is found that the suppression mechanisms of the instabilities are different between the current-driven type and the potential-driven type, which are selectively excited by the bias voltage applied to the small disk electrode.

In the case of the parallel flow velocity shear experiments, on the other hand, the ion-cyclotron instabilities are found to be excited by the parallel shear depending on the sign of the shear.
The fluctuation amplitude is observed to increase with increasing the shear strength, but the instability is found to be gradually stabilized when the shear strength exceeds the critical value.

Based on the results mentioned above, it is concluded that the plasma flow velocity shears in the magnetized plasma are important for controlling the ion-cyclotron instabilities, where the parallel and perpendicular shears play different roles in these instabilities, respectively.

\section*{Acknowledgment}

\hspace*{4ex}
We express our gratitude to Professor M. Koepke and Dr. G. Ganguli for their useful discussion and comments. We also thank Professor M. Inutake for his useful comments. The authors are indebted to H. Ishida for his technical assistance. This work was supported partly by a Grant-in-Aid for Scientific Research from the Ministry of Education, Culture, Sports, Science and Technology, Japan, and partly by the LHD Joint Planning Research program at National Institute for Fusion Science.

\end{document}